\documentclass[12pt,preprint]{aastex}
\begin{document}
\title{FORMATION OF NUCLEAR SPIRALS IN BARRED GALAXIES}
\author{H.B. ANN AND PARIJAT THAKUR}
\affil{Division of Science Education, Pusan National University,
 Busan 609-735, Korea}
\email{hbann@pusan.ac.kr, pthakur@pusan.ac.kr}
\begin{abstract}
We have performed smoothed particle
hydrodynamics (SPH) simulations for the response of the gaseous
disk to the imposed moderately strong
non-axisymmetric potentials. The model galaxies are composed of the three
stellar components (disk, bulge and bar) and two dark ones
(supermassive black hole and halo) whose gravitational potentials
are assumed to be invariant in time in the frame corotating with
 the bar. We found that the torques alone generated by the moderately strong
 bar that gives the maximum of tangential-to-radial force ratio as 
$(F_{Tan}/F_{Rad})_{max}= 0.3$ 
are not sufficient to drive the gas particles close to the center
due to the barrier imposed by the inner Lindblad resonances (ILRs). 
In order to transport the gas
particles towards the nucleus ($r<100$ \ pc), a central supermassive
black hole (SMBH) and high sound speed of the gas are required
to be present. The former is required to remove the inner inner 
Lindblad resonance (IILR) that prevents gas inflow close to the nucleus,
while the latter provides favourable conditions for the gas particles to
lose their angular momentum and to spiral in.
Our models that have no IILR show the trailing nuclear spirals
whose innermost parts reach close to the center in a curling
way when the gas sound speed is  $ c_{s} \gtrsim 15$ km s$^{-1}$. They
 resemble the symmetric two-armed nuclear spirals 
observed in the central kiloparsec of spiral galaxies. We found that 
the symmetric two-armed nuclear spirals are formed by 
the hydrodynamic spiral shocks caused by the gravitational torque of the bar 
in the presence of a central SMBH that can remove IILR when the sound 
speed of gas is high enough to drive a large amount of gas inflow deep 
inside the ILR. However, the detailed morphology of nuclear spirals
depends on the sound speed of gas.
\end{abstract}
\section{INTRODUCTION}
 Recent observations by ground-based telescopes with  adaptive optics and 
{\it Hubble Space Telescope (HST)} showed a variety of features
such as nuclear spirals,  nuclear rings, and nuclear bars in the
 centers of galaxies. Among them, nuclear spirals are most
 frequently observed ones in active galaxies \citep{reg99,mar99,pog02}
as well as in normal spirals galaxies \citep{phi96,lai99,lai01}.
The existence of these nuclear features had been suggested by 
the ground-based observation of the blue nuclei or peculiar nuclear 
structures since 1950s \citep{har56,ser58}, but analysis of their detailed 
morphology was impossible due to lack of resolution. However, thanks to the 
high resolution capability of $HST$, it is now possible to resolve the
nuclear features down to scales of $10$ pc or less  for the galaxies 
within $\sim 20$ Mpc.  

Just like the spiral arms in the disk of spiral galaxies, nuclear
 spirals have a diversity of morphology such as symmetric two-armed
 spirals, one-armed spirals, and chaotic ones \citep{mar03}. Because
 of the diverse morphology, the formation mechanism of
 nuclear spirals is difficult to be understood. However, some
mechanisms have been proposed to explain the formation of chaotic or
``flocculent'' and symmetric two-armed or ``grand-design'' nuclear
spirals. The flocculent nuclear spirals are suggested 
to be formed by the acoustic instability \citep{elm98,mon99},
whereas the grand-design nuclear spirals are shown to be generated by 
the gas density waves supported by the pressure forces in 
the non-self-gravitating gaseous disks \citep{eng00} or by 
the hydrodynamic shocks induced by the non-axisymmetric potential 
of the large scale bar \citep{mac02}. The weak nuclear spirals 
whose wave amplitude is considered to be linear can be explained 
by the density waves since the gas density waves can be applied 
to the regions where no shocks exist, while the strong nuclear spirals 
are thought to be the hydrodynamic shocks because their amplitude 
is likely to be out of the linear regime \citep{mac02}.

Since the morphology of nuclear spirals depends on the sound speed of gas 
and the shape of potential in the nuclear regions of 
galaxies \citep{eng00}, it is important to understand their effect on
the gas flows inside the ILR.  
The effect of sound speed on the gas flows in hydrodynamical simulations
is quite well known \citep{eng97,pat00,mac02}. However, the previous
studies except for that by \citet{mac02} mainly concerned with the 
gas flows outside the nuclear regions of galaxies. Using SPH
method, \citet{eng97} and \citet{pat00} showed that off-axis shock
developes at low sound speed, while on-axis one at high sound speed.
But the resolutions of their simulations are not high enough to analyze the
nuclear morphology in detail. 

The physical parameter that determines the shape of the gravitational 
potential in the nuclear regions of galaxies is the central mass 
concentration which can be much affected by the presence of an SMBH even 
though the mass of an SMBH is less than 1\% of the total mass of a galaxy. 
Since nuclear spirals are found to be preponderant in active
galaxies \citep{mar03} where SMBHs are expected to be present,
it seems to be important to understand the effect of 
the central  SMBH as well as that of the sound speed on the gas flows
inside the ILR. Thus, the purpose of the present paper is to see
the behaviour of the gas flows in the presence of the central SMBHs,
which leads to the formation of nuclear spirals. For this purpose,
we employ the SPH technique to solve the hydrodynamical equations for
the gas flow in the model galaxies with central SMBHs. We have also
included the self-gravity of the gas in our calculations, since
this can play significant role in the high density regions like shocks.

The remainder of this paper is organised as follows. In \S \ 2
 we describe the hydrodynamic modeling of the gas flows. In \S \ 3
 we present the results of our simulations and the discussion of our
results is given in \S \ 4. The conclusions are presented
in \S \ 5.
\section {HYDRODYNAMIC MODELING OF GAS FLOWS}
\subsection{Numerical Method}
We have used PMSPH method for modelling the gas flows in barred galaxies.
The hydrodynamical equations for the gas flow was solved by the smoothed 
particle hydrodynamics (SPH) method, while the self-gravity of gas was
calculated by particle-mesh (PM) N-body technique. In the SPH, 
fluid elements constituting the system are sampled and represented
by particles, and the dynamical equations are described by  
the Lagrangian form of the hydrodynamic conservation laws. Due to its 
Lagrangian nature, there is no grid to impose any artificial restrictions 
on the global geometry of the systems under study or any mesh-related 
limitations on the dynamic range in spatial resolution. 
As a result, the high numerical resolution is naturally achieved in high
density regions like shocks \citep[][]{fuk98},
which is necessary in our present problem of understanding the response
of gaseous disk in the nuclear regions of galaxies. This is the reason why
we have used the PMSPH code \citep{fux99,fux01}
which Roger Fux kindly made available. We describe here briefly 
the necessary specifications relevant to the particular aspects of our code
and will refer the readers to \citet{fux99,fux01} for the detailed
description of our code and to \citet{mon92} for general properties of 
solving the Euler's equation of motion using popular Lagrangian method. 

\subsubsection{Gas Dynamics}
The relevant form of the Euler equation of motions can be written as
\begin{equation}
\frac {d {\bf v}} {dt} \equiv \frac {\partial {\bf v}} {\partial t} +
  ( {\bf v}  \cdot {\bf \nabla})
 {\bf v} = - \frac {{\bf \nabla} (P+\Pi)} {\rho} - {\bf \nabla} \Phi \ ,
\end{equation}
where ${\bf v}({ \bf r})$ is the velocity field, and $\rho ({\bf r})$,
  $P({\bf r})$ and $\Pi({\bf r})$ are the density, pressure and
  artificial viscosity of the gas, respectively.  $-{\bf \nabla} \Phi$ is the
  gravitational force, where $\Phi$ includes
  potentials of the gas  as well as  the stellar and dark components.

In the SPH, the density is evaluated directly at each particle position
${\bf r}_{i}$ in space by summing the contributions from  the density
profiles of neighbouring particles and is represented  by
\begin{equation}
\rho_{i} \equiv \rho({\bf r}_{i}) = \sum_{j=1}^{N_{g}} m_{j} W({\bf r}_{i}
  - {\bf r}_{j}, h) \ ,
\end{equation}
where $N_{g}$ is the number of gas particles, $m_{j}$ is the mass of
individual particles, $W({\bf r}, h)$ is the kernel function and $h$ is the
smoothing length which defines the local spatial resolution and is
proportional to the local mean interparticle spacing. Here in our code 
the adopted SPH kernel is a spherically symmetric
spline that vanishes outside $2h$ because of its finite spatial extension. 
Thus, the particles only within a radius of $2h$  will contribute to 
the smoothed estimates for this kernel. 

Since we assume an isothermal gas, the equation of state is expressed
as
\begin{equation}
P_{i}= \frac {c^{2}_{s}} {\gamma}  \rho_{i} \ ,
\end{equation}
with the adiabatic index $\gamma = 1$ and $c_{s}$ as an effective sound
speed which can be interpreted globally as the velocity dispersion of
the interstellar clouds. The internal energy of the gas is at any time
and everywhere constant, and in particular, there is no energy equation
involved. The homogeneous velocity dispersion of warm gas component in
the external galaxies \citep[e.g.,][]{van84} partly justifies this
assumption. It means that cooling and heating exactly cancel each other
and the energy released by shocks is instantaneously radiated away.
With the isothermal assumption of the gas, the pressure gradient term
in the Euler's equation simply takes the form
\begin{equation}
\left(\frac {{\bf \nabla} P} {\rho} \right)_{i} \approx \frac {c_{s}^2}
 {\gamma}
  \sum_{j=1}^{N_{g}} m_{j} \left(\frac {1} {\rho_{i}} + \frac {1} {\rho_{j}}
  \right) {\bf \nabla}_{i}  W({\bf r}_{i} - {\bf r}_{j}, h), \
  \textrm{where $\gamma=1$}.
\end{equation}

For accurate treatment of flow near shocks, it is necessary to
  introduce the artificial viscosity. In our code, the artificial
 viscosity is calculated  as in \citet{ben90} with Balsara
(1995)'s correction, i.e.,
\begin{equation}
\left(\frac {{\bf \nabla} \Pi} {\rho} \right)_{i} \approx
  \sum_{j=1}^{N_{g}} m_{j} \widetilde{\Pi_{ij}} {\bf \nabla}_{i}
  W({\bf r}_{i} - {\bf r}_{j}, h) \ ,
\end{equation}
where
\begin{equation}
 \widetilde{\Pi_{ij}} \equiv \frac {\Pi_{ij}} {\rho^{2}}  =
  \left\{\begin{array}{ll}
 f_{ij} \frac {(-\alpha \mu_{ij} c_{s} + \beta \mu_{ij}^{2})} {\rho_{ij}}
 & \textrm{if $({\bf v}_{i}-{\bf v}_{j}) \cdot
({\bf r}_{i}- {\bf r}_{j}) \leq 0$}, \\ 0 & \textrm{otherwise},
\end{array} \right.
\end{equation}
\begin{equation}
\mu_{ij} = \frac {h ({\bf v}_{i}- {\bf v}_{j}) \cdot
  ({\bf r}_{i}-{\bf r}_{j})} {({\bf r}_{i}-{\bf r}_{j})^{2} + \xi h^{2}} \ ,
\end{equation}
with $\alpha=1.0$, $\beta=2.5$, $\xi = 0.01$ and $\rho_{ij} =
  (\rho_{i}+\rho_{j})/2$. Here, $f_{ij} = (f_{i} + f_{j})/2$ is
  the correction factor which was introduced by \citet{bal95} to avoid 
the energy dissipation in pure shearing flows and has the following form:
\begin{equation} 
f_{i} =  \frac {\vert \langle  {\bf \nabla} \cdot {\bf v}
   \rangle _{i} \vert} {\vert \langle  {\bf \nabla} \cdot {\bf v}
   \rangle _{i} \vert + \vert \langle  {\bf \nabla} \times {\bf v}
   \rangle _{i} \vert + 0.0001 c_{s}/h} \ .
\end{equation}
Furthermore, in above equation $(6)$, the term with the coefficient
 $\alpha$ corresponds to the bulk viscosity, whereas the term 
with the coefficient $\beta$ represents to the von Neumann-Richtmyer
 viscosity. The von Neumann-Richtmyer viscosity is necessary to handle
 strong shocks. On the other hand, the bulk viscosity is required to
 damp post-shock oscillations.

In order to achieve high-resolution, it is necessary to treat smoothing
length $h$ properly since it determines local spatial resolution.
Basically, the SPH assumes that $h$ is the same for all the particles, which 
would give relatively more accurate estimates in the high density 
regions than in the lower ones. Thus, it is  desirable  
to compute smoothed quantities with the same level of accuracy everywhere
in order to achieve both consistency and efficiency in 
the smoothed estimates. For this purpose, our code considers 
spatially variable smoothing length $h$ to assign 
an individual smoothing length $h_{i}$ to each particle in such a manner
that the  number of neighbour particles $N_{i}$ to each particle 
always remains as close as possible  to a fixed number $N_{o}=35$.
This allows us to save computing time significantly while 
keeping spatial resolutions high enough to resolve the shocks 
\citep[][]{fux99,fux01}. In this scheme, two particles $i$ and $j$
 are defined as mutual neighbours 
 if $i \ne j$ and $\vert {\bf r}_{i} - {\bf r}_{j} \vert < 2 h_{ij}$,
 where $h_{ij} =  (h_{i} + h_{j})/2$ is the symmetrized smoothing
 length which is supplied  in all the above formulae for pressure gradient
 and artificial viscosity to ensure momentum conservation as well as
 to treat shock phenomena in a gaseous disk correctly. In our code,
the update of the smoothing lengths was made at each time step according
 to a general three-dimensional scaling law:
\begin{equation}
\frac {h_{i}} {h_{o}} = \left(\frac {N_{i} +1} {N_{o} +1} \frac
  {\rho_{o}} {\rho_{i}} \right)^{1/3} \ ,
\end{equation}
where ${h_{o}}$ and  ${\rho_{o}}$ are constants, $N_{i}$ is number of
neighbours of particle $i$, and ${+1}$ is added to take into account
particle $i$. Since the determination of the density $\rho_{i}$ at the
particle location requires the $h_{i}$'s which are not known a priori, it
is better to take the time derivative of equation $(9)$ and substitute
the continuity equation to yield  
\begin{equation}
\dot{h_{i}} \equiv \frac {d h_{i}} {d t} = \frac {1} {3} h_{i}
  \left(\frac {1} {N_{i}+1} \frac {d N_{i}} {d t} + \lbrack
  {\bf \nabla} \cdot {\bf v} \rbrack_{i} \right) \ .
\end{equation}
In the traditional approach of \citet{ben90}, the term ${d N_{i}}/{d
t}$ was set to zero to ensure a constant number of neighbours. However,
this does not prevent a slow numerical departure of the $N_{i}$'s
from $N_{o}$ with time. Instead, our code takes the advantage of the
 $N_{i}$ term to damp such departures  by setting
\begin{equation}
\frac {d N_{i}} {d t} = \frac {N_{i} - N_{o}} {\eta \triangle t}
\end{equation}
in equation $(10)$ and integrates the resulting equation along with the
 equations of motion. The parameter $\eta$ is used to control the
 damping rate per time step $\triangle t$ and should be significantly
 greater than $1$ to avoid sharp changes in non-gravitational forces. 
 We have fixed  $\eta = 5$ in all our simulations.

Since the computation time depends much on the smoothing lengths of
individual particles, it is customary to impose quite a large lower limit 
on the minimum smoothing length $h_{min}$ to save computing time.
However, we have used $h_{min}$ =0.1 pc to resolve structures
down to pc scale. Our adopted $h_{min}$ is an order of magnitude smaller 
than the $h_{min}$'s employed in the previous works \citep{hel94,fuk00}.
Imposing virtually no limit on the  $h_{min}$ was possible due to the
fast neighbour searching algorithm adopted in our code.
However, the smallest smoothing length actually achieved in our simulations
is $\sim 2$ pc which is still smaller than the $h_{min}=5$ pc used in
the high-resolution simulations of \citet{fuk00}.
Another important aspect of our code which is worth to be noted
 here is the adoption of a synchronised version of the standard leap-frog 
time integrator \citep[e.g.,][]{hut95}. This integrator is known to
 well conserve the total energy and the total angular momentum and 
also incorporates an adaptative time-step to temporally resolve 
the high density shocks in the gaseous disk \citep[see][]{fux99,fux01}.

\subsubsection {Self-gravity of Gas}
Since the mass of the gaseous disk is assumed to be only a tiny
fraction of the model galaxy $(\sim 1\%)$, our gaseous disk is not a 
self-gravitating one. But, we have included the self-gravity of gas
in our calculations because this can play significant role in the 
high density regions such as shocks. In particular, the effect of
gas self-gravity can not be ignored in the later evolution of gaseous
disk, especially in the nuclear disk where the nuclear spirals are 
supposed to be formed.  For computing the  
self-gravity of gas, we have used particle-mesh (PM) technique
with polar-cylindrical grid geometry and variable homogeneous ellipsoidal
kernel for the softening of the short range forces \citep{pfe93}.
Since our simulations are confined in two-dimensions to achieve a
high spatial resolution, we have used a single grid
version of PMSPH code of \citet{fux99,fux01}. For more detailed description of 
the PM technique, we simply refer the readers to \citet{fux99,fux01} 
which itself follows \citet{pfe93} for this technique. 
Owing to the intrinsic nature of polar-cylindrical grid, the
radial and azimuthal resolutions are increased toward the center where a
variety of features are likely to be developed due to the gas 
inflow driven by the bar.

\subsection{The Model Galaxy}

We assume that a model galaxy is composed of the three stellar
components (bulge, disk and bar) and two dark components (SMBH and
halo). The properties (density, size and structure) of all these
potential generating components are assumed to be invariant in time in
the frame corotating with the bar. We employ the same analytic
potentials as those of \citet{ann00}  and \citet{ann01} for the stellar
components and the halo. 
  
\subsubsection{The Potential}

The halo component, which gives rise to the flat rotation curve at
outer radii, is assumed to have a logarithmic potential with finite
core radius,
\begin{equation}
\Phi (r)_{halo} = {1\over 2} v_0^2 \ln(R_h^2+r^2) + const,
\end{equation}
where $R_h$ is the halo core radius, and $v_0$ is the constant
rotation velocity at large $r$. For the disk component, we adopt the
 \citet{fre70} exponential disk which has the following form
 for the potential,
\begin{eqnarray}
\nonumber \Phi (r)_{disk} &=& \pi G \Sigma _0 r \left[ I_0\left(r
\over 2R_d\right)
K_1\left(r \over 2R_d\right)\right. \\
&-&\left. I_1\left(r \over 2R_d\right)K_0\left(r
\over 2R_d\right)\right].
\end{eqnarray}
Here $R_d$ is the disk scale length, $\Sigma _0$ is the central surface
density, and $I_0,~K_0, ~I_1$, and $K_1$
are the modified Bessel functions. The total disk mass is
simply $M_{disk}=2\pi \Sigma_0 R_d^2$. As for the bulge component,
 we have assumed the Plummer  model \citep{bin87}
\begin{equation}
\Phi (r)_{bulge} = - {G M_{bulge}\over \sqrt {r^2+R_b^2}},
\end{equation}
where $R_b$ is a parameter which controls the size of bulge. The bulges of
real galaxies can be better represented by de Vaucouleurs' $R^{1/4}$-law,
but we have taken Plummer model for simplicity.

The bar is a triaxial component in three-dimension. However, since our
 simulation is restricted to the two-dimensional disk, we have
 adopted the following form of gravitational potential proposed
 by \citet{lon92} for the bar component,
\begin{equation}
\Phi (r)_{bar} = {G M_{bar}\over 2a}
 \log\left({x-a+T_- \over x+a+T_+}\right),
\end{equation}
where $T_\pm=\sqrt{[(a\pm x)^2+y^2+b^2]}$. Here $a$ and $b$ are
 the needle length and softening length, respectively. 
 Because \citet{lon92}'s potential is of a softened needle, it 
does not describe an axisymmetric system for $a=b$.
 The parameter $a/b$ defines the elongation of the bar which is
 proportional to the axis ratio $\tilde{a}/\tilde{b}$ of  
\citet{fre66} flattened
 homogeneous ellipsoid bar
 when $a/b \ge 2$ \citep[see Fig. 3 of][]{lon92}, where $\tilde{a}$
 and $\tilde{b}$ are the major and minor axes of flattened homogeneous
 ellipsoid bar. After readjusting the relations established in \citet{lon92}
 for the parameters of
these two bars, we found that
$\tilde{a}/\tilde{b} = 1.05 \ a/b$  with $\tilde{a}= 1.15 \ {a}$
 and  $\tilde{b}= 1.1 \ {b}$.

We assume the point mass potential for the central SMBH component,

\begin{equation}
\Phi (r)_{SMBH} = - {G M_{SMBH}\over \sqrt {r^2+\epsilon_{SMBH}^2}},
\end{equation}
where $M_{SMBH}$ is the mass of the central SMBH and $\epsilon_{SMBH}$ is
 a softening parameter.

\subsubsection{Set up of the Gaseous Disk}

We assumed an isothermal uniform gaseous disk with
an infinitesimal thickness. Since our aim is to simulate the response of the
gaseous disk in early type galaxies,  the mass of gaseous disk was chosen to be
1\% of the total visible mass of the model galaxy. To make an initial uniform
gaseous disk, we distributed $2 \times 10^4$ SPH particles randomly within the
radius of $5$ kpc with the initial rotational velocities balancing
to the centrifugal acceleration caused by the axisymmetric
components. The initial resolution length, defined by
$\sqrt{m_{i}/ 4 \pi \Sigma(R_{g})}$, was found to be  $\sim20$ pc,
where $m_{i}$ and  $\Sigma(R_{g})$ are the mass of each gas particle
(in the unit of $M_{sc} = 2 \times 10^{11} \rm M_{\odot}$) and
the surface density of the initial gaseous disk of radius $R_{g} = 5$ kpc
(in the unit of $\Sigma_{sc}= 2 \times 10^{5} \rm M_{\odot}$ pc$^{-2}$),
 respectively. The evolution of gaseous disk is turned on when we introduce
the non-axisymmetric potential of the bar. But the strength of bar is
 increased gradually within a half bar rotation period  to avoid spurious
response of the gaseous disk, where bar rotation period
 $\tau_{bar}$ is $\sim 1.4\times 10^8$ yr.
\subsubsection {Model Parameters}
\clearpage
\begin{table*}
\begin{center}
{TABLE 1}\\
\vskip 0.2cm
{\small MODEL PARAMETERS}
\vskip 0.2cm
\begin{tabular}{ccc}\hline \hline
Model &  $c_{s}$ [km s$^{-1}$]  & $M_{SMBH}$  \\ \hline
M1  &    10. &  0.0     \\ 
M2  &    10. &  0.002   \\
M3  &    15. &  0.002   \\
M4  &    20. &  0.002   \\
\hline
\end{tabular}
\end{center}
\end{table*}
\clearpage
We have considered four models M1-M4 which have the same mass models
except for the presence of the central SMBH but different 
hydrodynamic properties. Basically the mass distributions of the 
model galaxies are assumed to be similar to that of an early type barred
galaxy ($\sim$ SBa) of which disk-to-bulge mass ratio (D/B) is 2 and the 
fractional mass of the bar $M_{bar}/M_{G}$ is 0.2. In order to reproduce
 the typical rotation curves of early type barred galaxies, the scale lengths
 of bulge, disk, bar and halo are chosen to be $R_{b}=0.5$ kpc,
 $R_{d}=3.0$ kpc, $a=3.0$ kpc and $R_{h}=15.0$ kpc, respectively. 
Since the total visible mass of the
model galaxies $M_{G}$ is assumed to be $\sim 4\times 10^{10} \rm M_\odot$,
the masses of the gas, bulge, disk and bar components in the unit of
  $M_{sc} = 2 \times 10^{11} \rm M_{\odot}$ are $M_{gas}=0.002$,
$M_{bulge}=0.054$, $M_{disk}=0.104$ and $M_{bar}=0.04$, respectively. 
 Furthermore, the rotation velocity of halo and the SMBH softening parameter
  are fixed at $v_{0}=186$ km s$^{-1}$  and
  $\epsilon_{SMBH} = 1$ pc, respectively. We chose moderately 
strong bars that gives the maximum of tangential-to-radial 
 force ratio as $(F_{Tan}/F_{Rad})_{max}=0.3$ by assuming $a/b=3$
$a/b=3$ with $M_{bar}/M_{G}=0.2$ for all the models. In the models with
a central SMBH, we assumed the mass of an SMBH ($M_{SMBH}$)
to be $\sim 1\%$ of the total visible mass of host galaxy
to remove the IILR effectively.
The fractional mass of the SMBH assumed in the present models is somewhat
larger than the observed one \citep[e.g.,][]{marc03}. But, we take it
for an easy comparison with the previous works \citep[e.g.,][]{fuk00}.
Since the primary role of the SMBH in our models is to remove the IILR
and there is not much difference in the gas flows in models with different
SMBHs unless they are too small to remove the IILR \citep{ann04}, it should
give us insight into the inner dynamics of galaxies that might host SMBHs.
We list all the models in Table 1
where $c_{s}$ and $M_{SMBH}$ stand for the effective sound speed of
 the gas and the mass of SMBH 
in the unit of $M_{sc} = 2 \times 10^{11} \rm M_{\odot}$, respectively. 

As clearly seen in Table 1, the model M1 does not have
 a central SMBH component, 
while the other models contain SMBHs at the center. Hence, the model M1
was selected here to compare the effect of SMBH on the gas flow, 
 whereas the models M2-M4 were chosen to see the effect of sound speed in 
the presence of an SMBH. According to the terminology of \citet{mac03},
 the models M1-M2 that assume $c_{s}=10$ km s$^{-1}$ are the 
cold gas models and  remaining two models (i.e., the models M3-M4) 
can be treated as the hot gas models. 
 Since the models M2-M4 have the same mass distribution, we present the 
rotation curves and angular frequencies curves of M1 and M2 in Fig. 1.
The top left panel of Fig. $1$ represents the rotational velocities of
the bulge, disk, bar and halo components of the model M1 as a function
of radius, while top right panel shows rotational curves of the model M2
which includes the contribution of the central SMBH besides all other
components those in the model M1.
Due to non-axisymmetric nature of the bar, its contribution to
the rotational velocity is included here after averaging 
 it axisymmetrically \citep{has93}. As can be seen clearly in 
 the top panels of Fig. 1, the presence of SMBH does not much affect 
the rotational velocities except for the very vicinity of galactic nuclei.
However, as evident from the bottom panels of Fig. 1 where we present 
the angular frequencies corresponding to the rotational velocity generated
by the axisymmetric components as a function of radius, the existence of 
the IILR critically depends on 
the presence of SMBH.
For the adopted bar pattern speed of 
$\Omega_{p}$=44.6 km s$^{-1}$ kpc$^{-1}$, which makes 
the corotation radius ($R_{CR}$) similar to $1.2 a$, we have two ILRs
(i.e., the IILR and the outer inner Lindblad resonance (OILR)) 
in the model M1 but no IILR in the model M2. The locations of IILR and 
OILR are, respectively, found at $0.3$ kpc and $1.5$ kpc in the model M1, 
whereas the location of the single ILR in the model M2 is the same as 
that of the OILR in the model M1.

\clearpage

\begin{table*}
 \centering
{TABLE 2.} \\
\vskip 0.2cm
{\small AVERAGE PITCH ANGLE FOR NUCLEAR SPIRALS}
\vskip 0.2cm
\begin{tabular}{cccccccc}  \hline \hline
{Model}  & $c_{s}$ [km s$^{-1}$]  & \multicolumn{5}{c}
 {$\overline{i_{p}}$ [degree]}\\
      & &   & $5 \tau_{bar}$ & & $10 \tau_{bar}$ & $20 \tau_{bar}$ \\ \hline
M2 & 10  &   & 12.8  & & 10.5 & 9.6 \\
M3 & 15  &   & 13.2 & & 12.5 & 11.8 \\
M4 & 20  &  & 14.5 & & 13.4 & 12.5\\
\hline
\end{tabular}
\end{table*}

\clearpage

\section {RESULTS}

\subsection {Time Evolution of the Gaseous Disk}
\subsubsection{Models with SMBH}

Since the global morphological evolutions of the gaseous disk of our models 
are similar, we present this only for the model M3 which considers the medium
value of the sound speed (i.e., $c_{s}=$ 15 km s$^{-1}$) employed in the
 present models. However, the evolution of the nuclear regions
  are presented for all the models to understand the behaviour of 
 the gas flow in the central kilo-parsecs of barred galaxies. Here
 our results are shown on the frame co-rotating with the bar
 which always lies horizontally.
  Fig. 2 shows the global distribution of the gas particles in the
 model M3 at the four evolution times that are given in the upper left corner
 of each panel. At time of $0.75 \tau_{bar}$, the gaseous disk is
 distorted in the direction leading the bar major axis 
 by $\sim 135^{\circ}$, displaying the spiral shocks
such as the spiral shocks in the outer disk,
4/1-spiral shocks just inside the bar radius \citep{eng97},
and the principal shocks along the leading edges of the bar which are almost
straight, off-centered and inclined to the bar major axis. These shocks 
are generic features for various bar potentials and are independent of
sound speed in the gas \citep{mac02}.
Moreover, they are made by the gas flows driven by the gravitational
torques of the bar and reflect the orbital families in the underlying 
potential. As shown in the velocity field of the model M3 at the time
 of $0.75 \tau_{bar}$ in Fig. 3, the gas particles move inward alongside
 the principal shocks following the x1-orbits and then swing around near 
 the ILR towards the far side of the symmetric shocks.
Some of the particles move to the x2-orbits inside the ILR and others move 
outward reaching the far side-shock again. The winding inner part of 
the principal shocks inside
the ILR are made by this process of orbital switching from the x1-orbits 
which align with the bar axis to the x2-orbits perpendicular to 
the bar \citep{ath92, wad94,kna95,eng00}. 

At the time of $5 \tau_{bar}$, the three above mentioned
spiral shocks become less prominent, but we can see 
the trailing spiral arms in the outer disk.
The nuclear region inside the ILR looks like oval disk with a hole
in the center which seems to lead the bar by more than  $45^{\circ}$.
The reason for the weakening of the shock features are
due to the reduced gas flows arising from the lack of gas recycling in
our models. In the later evolution times,
shocks become very weak, since most of the particles inside
the bar radius have already  moved into the central kiloparsec.
This makes the completely filled nuclear disk. 
Furthermore, it also seems that the outer spiral arms get disappeared
and evolve to more symmetric structure. Since we do not
employ gas recycling, our models do not reach to the steady state even at 
the time of $\sim 20 \tau_{bar}$. However, the global morphology of
the gaseous disk does not change much and it reaches 
to a quasi-steady state after $\sim 20 \tau_{bar}$.

Fig. 4 shows the evolution of gas distribution in the central kiloparsecs 
of the model M3. Now we can clearly see a two-armed spiral in
the nuclear disk. The early phase of evolution (i.e., at $0.75 \tau_{bar}$), 
shown in the top left panel of Fig. 4, is characterized by 
a development of open symmetric spiral patterns that are connected 
to the principal shocks in the leading edge of the bar. These spiral
patterns display clumpy substructures that are similar to the ripples and clumps
observed in the recent numerical simulations for spiral shocks \citep{wad04}
which suggest that these substructures are made by the wiggle instability.
However, this clumpy structure disappears shortly within $\sim \tau_{bar}$ and 
a smooth two-armed nuclear spiral pattern begins to emerge. 

At the time of $5 \tau_{bar}$, the trailing nuclear spiral arms are well
developed and they extend more than ${3 \over 2} \pi$ in the azimuthal
direction but the density of the spiral arms decreases significantly after
${3 \over 2} \pi$. In the later evolution times, the nuclear spiral becomes
more tightly wound and its innermost parts reach closer
to the center in a curling manner. This results in the increase of 
winding angle of nuclear spiral which becomes greater than $3 \pi$
at the time of $20 \tau_{bar}$.
At this time, there are much more particles in the nuclear disk but 
the arm-interarm density contrast becomes smaller than earlier ones. 

In order to see the effect of the sound speed on the gas flows
in the nuclear regions of galaxies, we present the evolutions of the 
central kiloparsec of the models M2 and M4 in Fig. 5.
Despites the similarity in  the global morphological evolutions of
the models M2 and M4, the morphological evolutions of the central kiloparsec
of the two models, both of which show the formation of nuclear spirals,
are quite different. Besides the morphological difference of 
nuclear spirals, the evolution of the model M4 that assumes
$c_{s}=$ 20 km s$^{-1}$ is much faster than the model M2
whose sound speed is half of the model M4. As a result,
nuclear spirals are formed in the earlier time for the models
with higher sound speed than those with lower sound speed.

However, the earlier evolutions of the nuclear regions of the gaseous
disks with different sound speeds are not much different. They are 
characterized by the open spiral patterns of clumpy substructures, as similar
to those of the model M3. The drastic changes have been noticed at the 
later evolutions. At the time of $5 \tau_{bar}$, the model M2 shows
 rather tightly wound symmetric two-armed nuclear spiral whose winding angle
  is less than ${3 \over 2} \pi$, while the model M4 shows 
 spiral pattern similar to that of the model M3 but extends more
closer to the center, making its winding angle greater than 
$ 2 \pi$.
It is also apparent that the models with higher sound speeds have
larger number of the gas particles in their nuclear disks. 
But the arm-interarm density contrast is highest in the models with
 the lowest sound speed. At this time, the density contrasts between
 arm and interarm regions are greater than those of
 the principal shocks due to the significant reduction of the 
gas flows along the bar. 

At the time of $10 \tau_{bar}$, the nuclear disk of the model M4 shows
a well developed symmetric two-armed nuclear spiral whose innermost parts
reach very close to the center to wind it upto $\sim 3 \pi$, 
whereas that of the M2 shows a symmetric two-armed spiral whose 
density decreases significantly after $\sim {3 \over 2} \pi$ 
in the azimuthal angle. In the later 
evolution time ($\sim 20 \tau_{bar}$), the innermost parts of 
nuclear spiral arms of the model M4 are no longer resolved after $2 \pi$
 due to the formation of clumps close to the center, 
while the nuclear spiral of the model M2 evolves to 
 more tightly wound ring-like spiral whose winding angle is appeared 
to be more than $2 \pi$. Also, there seems to be little gas in
 the central hundred parsecs of the model M2. The diameter of 
the nuclear disk in quasi-steady state evolution also seems to 
depend on the sound speed of gas. The models with lower sound speeds
allow larger nuclear disk than those with high sound speed.

The open spiral is defined by the larger pitch angle $i_{p}$, while relatively
 smaller pitch angle $i_{p}$ represents tightly wound one. 
 Since the pitch angles $i_{p}$ of the above mentioned symmetric two-armed
 nuclear spirals do not vary much along the radius, we have measured
 the mean pitch angle $\overline{i_{p}}$ in order to define their openness.
 Table $2$ summarizes the average pitch angles 
$\overline{i_{p}}$ of nuclear spirals at the three evolution times of
 $5 \tau_{bar}$, $10 \tau_{bar}$ 
 and $20 \tau_{bar}$ for the models M2-M4. As can be seen clearly 
in Table 2, the average pitch angle $\overline{i_{p}}$ increases with 
the sound speed which suggests that the nuclear spiral arms open out as
sound speed in the gas increases. Furthermore, it is also apparent that
the symmetric two-armed nuclear spirals become slightly more 
tightly wound with time. 

\subsubsection{A Model without an SMBH}

The evolution of the central kiloparsec of the model M1 is much different 
from the other models. However, the global morphological evolution 
of the model M1 is nearly similar to that of the model M2 which has 
identical initial conditions as those in the model M1 except for 
the presence of an SMBH. The drastic difference in the nuclear evolution 
is due to the presence of the IILR in the model M1 that can prevent 
the inflow of gas toward the nucleus. Fig. 6 shows the evolutions of 
the central kiloparsec of the model M1 at six evolution times
which are indicated by the numbers in the upper left corner of 
each frame. The early evolution at $0.75 \tau_{bar}$ is similar to 
that of the model M2 but the snapshot at $5 \tau_{bar}$ already 
shows much difference. There is a development of a ring inside 
the two symmetric spiral arms near $\sim$ $1$ kpc. This ring is 
located near the IILR
and it becomes more elongated until $\sim 12 \tau_{bar}$, resulting in the 
development of the highly elongated oval-shaped nuclear ring 
with its orientation close to the bar minor axis. 
At the time of $20 \tau_{bar}$,
the eccentricity of the oval-shaped nuclear ring decreases
and some of the gas particles around the IILR acquire angular momentum
to come out from this resonance due to the positive 
bar torque there \citep{com02}. As a result, nearly circular nuclear 
ring-like 
structure with a large diffusion of the gas particle around its boundary is found.

The later evolutions (from $\sim 10\tau_{bar}$) of the M1 model lead to the
formation of a pair of leading nuclear spirals that emerge from 
the nuclear ring structure
just outside the IILR. It reaches to the quasi-steady state at 
$\sim25\tau_{bar}$. There seems to be only a negligible number of the 
 gas particles inside the IILR due to the change in the sense of 
the bar torque at the IILR \citep{com02}. Similar morphologies were 
also observed in the simulations of \citet{hel01} who called 
the two-armed spiral and ring structure as a double rings before
they merger into an elongated nuclear ring. But, the leading nuclear
spirals are 
transient features in their models and the elongated nuclear ring persists 
for the end of run with varying orientation. The reason for this discrepancy in 
the persistence of the leading nuclear spirals is not clear. It might
be due to the differences in the mass models as well as the
neglection of gas self-gravity in their models.
However, there seems to be not much difference in the potential shapes 
which directly affect the gas flows, since
the evolution until about $\sim 10 \tau_{bar}$ is virtually identical.
Thus, it seems quite plausible that the difference in the later evolution
is due to the effect of gas self-gravity on the gas flow. In order 
to see whether the gas self-gravity is the real cause for
the discrepancies between our models and their ones, we performed 
a simulation identical to the model M1 except for the neglection of
gas self-gravity, which shows the transient leading nuclear spirals similar
 to those of \citet{hel01}. Moreover, the evolution of the elongated 
nuclear ring 
after the formation of the transient leading nuclear spirals, 
which is characterized
by the highly elongated shape and continuous change in the orientation,
is also observed in this case.
This suggests that the self-gravity of gas help to sustain 
the leading nuclear spirals by enhancing density contrast. 
However, the different treatment of hydrodynamical parameters 
such as artificial viscosity and limiting smoothing length ($h_{min}$), 
along with the different assumption of the mass models may also 
play some roles in making the discrepancies between our model M1 
and those of \citet{hel01}.

\subsection {Bar-driven Nuclear Spiral Shocks} 

As described in the previous sections, all the models in the present 
study show a development of well defined nuclear spiral in their quasi-steady
state evolution. Since these nuclear spirals seem to be connected with the
principal shocks which are thought to be genuine hydrodynamical 
shocks \citep{ath92,eng97,mac02}, it is quite plausible that, like 
the principal shocks, they are also formed by the hydrodynamical shocks due to 
 the gas flow driven by the gravitational torques of the bar. 
\citet{mac02} noticed that strong spiral shocks comparable to the principal 
shocks can propagate deep inside the IILR to make nuclear spiral
when the sound speed of gas is high ($c_{s}=20$ km s$^{-1}$), 
while  for the low sound speed of gas ($c_{s}=5$ km s$^{-1}$), 
the principal shocks can not penetrate the IILR barrier. Since
our models except for the model M1 have no IILR, we expect that the
trailing nuclear spiral patterns revealed in the models M2-M4 are
formed by the spiral shocks that penetrate deep inside the ILR.

In order to see whether the nuclear spiral patterns shown in Fig. 4 are made of 
shocked gas particles, we examine the global distribution of gas particles 
which are shocked by supersonic flow, i.e.,
$-h {\bf \nabla} \cdot {\bf v} > c_{s}$ \citep{eng97}, for the
model M3 at the evolution time of $5 \tau_{bar}$ in Fig. 7. 
As is expected, 
the regions associated with the well known spiral shocks
such as the principal shocks and 4/1-spiral shocks are 
well populated by these highly shocked gas particles. Thus, we believe
that the nuclear spirals developed in the model M3 are formed by hydrodynamical
shocks driven by the bar. 
However, there are some differences between the nuclear spiral pattern 
revealed by the distribution of all the gas particles in the central 
kiloparsec, as shown in Fig. 4, and that of the shocked gas particles in Fig. 7. 
The nuclear spiral shown in Fig. 4 displays a symmetric two-armed spiral
pattern that winds up less than ${3 \over 2} \pi$ in the azimuthal angle with
similar density over the whole extension, 
whereas the spiral pattern made by shocked gas particles 
in Fig. 7 shows a somewhat sudden density drop near the azimuthal angle
of $\sim {3 \over 4} \pi$ but it extends upto ${2} \pi$.
There is a slight mismatch between the locations of these 
two spiral patterns. The loci of the local density maximum representing 
the nuclear spiral pattern in Fig. 4 are located just  outside
the nuclear spiral pattern revealed by shocked gas particles.
But, this mismatch between the locations of the two spiral
patterns is consistent with the general pattern of density distributions
in the shocked regions. The loci of the highly shocked gas particles
represent the shock front,
while the regions of density maximum represented by the nuclear spiral 
pattern in Fig. 4 are located in the post-shock region.
The density drop in the spiral pattern of shocked gas particles seems
to be caused by the process of orbital switching from the x1-orbits 
to the x2-orbits.

Fig. 8 displays the distribution of shocked gas particles in the central 
kiloparsec of the models M1-M4. We selected earlier 
evolution time (i.e., at $3 \tau_{bar}$) for the models M3 and M4, since 
the models with high sound speed evolve faster than the low sound speed models.
 As can be seen in Fig. 8, the distributions of shocked gas particles 
in the central kiloparsec of all the selected models reveal well defined 
spiral patterns which look very similar to the nuclear spiral shock pattern
 seen in Fig. 7. All the models show a smooth connection of 
nuclear spiral shocks with the principal shocks and 
a drop in the arm amplitude near the azimuthal angle of 
${3 \over 4} \pi$. But, for the models M1-M4, they extend upto 
winding angles of less than  ${3 \over 2} \pi$, $\sim {3 \over 2} \pi$, 
$\sim 2 \pi$ and more than $2 \pi$, respectively.
 Thus, it seems apparent that the
symmetric two-armed nuclear spirals can be formed by the bar-driven
spiral shocks, regardless of the sound speed in the gas and the presence
or absence of the central SMBH. However, the trailing nuclear spiral 
shocks can not be strong enough to maintain the nuclear spiral pattern in 
the later evolutions when regions of the principal shock are cleared
of gas. Since the devoid of gas in regions of the principal shock 
is due to the lack of gas recycling in our models, we expect that 
the bar-driven spiral shocks can maintain the nuclear spiral arms even in
the later evolutions if we supply the gas continuously.
In case of the model M4 
which shows the fastest evolution, only a small fraction of highly shocked
particles can penetrate deep inside the ILR after $\sim 3 \tau_{bar}$ to make
nuclear spiral arms. But, this weakening of the trailing nuclear spiral shocks
occurs at $\sim 7 \tau_{bar}$ for the model M2, since it shows 
the slowest evolution. 
For the model M1 which has an IILR at $\sim 0.3$ kpc,
the trailing nuclear spiral shocks disappear after $\sim 10 \tau_{bar}$ 
due to the development of
leading nuclear spiral shocks propagating outward from the IILR where
the infalled gas particles are accumulated due to the barrier caused by this
resonance and acquire angular momentum from the bar. 

\subsection {Gas Inflow} 

As shown in Fig. $2$, the evolution of the gaseous disk leads
to the redistribution of the gas particles. Although some fractions
of the gas particles move outward by gaining angular momentum from the
bar, a larger fractions of the gas particles move inward due to the loss
of angular momentum. Since nuclear features are formed by the gas that
moves inward across the ILR, it is 
important to understand the gas inflow in the central kiloparsec in detail.
Moreover, the gas inflow deep inside the ILR may be closely related
to the fueling mechanism of AGNs. 
Fig. 9 presents the time evolution of the fraction of particles 
accumulating inside two radial zones of $1.5$ kpc and $0.2$ kpc. 
Since the ILR (the OILR for the model M1) is located near 1.5 kpc, 
 the gas accumulated inside the radial zone of 1.5 kpc shown in 
 the left panel of Fig. 9 reflects the gas inflow across the ILR 
 which is characterized by a large amount of gas inflow in 
the early phase of evolution with little difference among the models. 

However, there are remarkable differences in the gas inflow close to
the nucleus. As can be inferred from the time evolution of the
fraction of gas particles accumulating inside the inner $0.2$ kpc of the 
modeled gaseous disks shown in the right panel of Fig. 9, 
the models M1 and M2 have
negligible gas inflow into the central hundred parsecs except for 
some inflow around $\sim 10 \tau_{bar}$ in the model M1, 
 while the models M3 and M4 show appreciable inflow throughout 
 the evolution. Furthermore, it is also apparent that the gas inflow rate
  increases with the sound speed for the models with the central SMBH. 
 As a result, the highest gas inflow rate is found for the model M4,
  which  is a factor of $\sim 5$ larger than that of the model M3. 
The correlation between the amount of gas inflow close to the center 
and the gas sound speed is similar to that found by  
 \citet{pat00} although their models do not take into account 
the central SMBHs. Apart from this consistency, it is also important
 to note that due to lack of resolution, the models considered 
 by \citet{pat00}, comparable to our model M1, do not develop 
a nuclear leading spiral. However, they show much larger inflow towards 
the center as sound speed increases.

The different amount of gas inflow among the models with the same 
 potential but different sound speeds implies
that angular momentum evolution depends on the sound speed in the gas,
since the gas particles can move inward only if they lose their angular 
momentum. Fig. 10 shows the dependence of angular momentum 
evolution on the sound speed in the gas. The total angular momentum 
inside the bar radius (i.e., inside $r = 3$ kpc) where the inflow of gas is 
taking place decreases
fastest in the model M4 which has the highest sound speed in the gas 
(i.e., $c_{s}=20$ km s$^{-1}$),
while it decreases slowest in the model M2 with $c_{s}=10$ km s$^{-1}$.
Thus, it suggests for the models with the central SMBH that 
the gas particles in the high sound speed mediums lose their 
angular momentum more easily than those in the low sound speed mediums.
The correlation between the rate of angular momentum loss 
and the sound speed in the gas can be understood if we consider the
fact that the viscosity of the hot gas is greater than that of the cold gas
(see Eq. (6)). Besides this dependence of angular momentum loss on 
the sound speed for the models having an SMBH, it is also apparent 
that the presence of central SMBH much affects the evolution of 
total angular momentum inside the bar radius in the case of low sound speed 
medium (i.e., $c_{s}=10$ km s$^{-1}$), since it decreases faster 
in the model M1 than in the model M2. This suggests that the gas particles 
in the cold gas models can lose their angular momentum relatively easily 
for the model without an SMBH (i.e., the model M1) than 
that with an SMBH (i.e., the model M2).

There is some difference in the gas inflow between the models M1 and M2
that assume the same sound speed but different potentials due to 
the absence or the presence of an SMBH.
The larger amount of gas inflow in the model M1 than in the model M2 around 
$\sim 10 \tau_{bar}$ 
is due to the larger $F_{Tan}/F_{Rad}$ in the central kiloparsec of the
 model M1, since the model M1 does not consider the central SMBH. 
The abrupt drop after $\sim 10 \tau_{bar}$ of the fraction of gas 
particles accumulated inside
 the central $0.2$ kpc of the model M1 is resulted from the onset 
of the outflow at the IILR which is caused by the change in
 the sense of bar torques at this resonance \citep{com02}. 
Particles gain angular momentum at the IILR and can move outward
from there. This outward gas flow leads to the formation of 
leading nuclear spiral in later times. 

It is interesting to see whether the gas inflow of our models can 
supply enough fuels for the  AGNs. To do this, we estimated the mean inflow rates
across the radius of $0.2$ kpc which is inside the
IILR of the model M1. But as expected from Fig. 9, the mean inflow rate of the
model M1 and M2 are negligibly small and can not fuel the AGNs.
However, the mean inflow rates within the central $0.2$ kpc of the models 
M3 and M4 are $0.004$ $\rm M_\odot$ yr$^{-1}$ and $0.013$ $\rm M_\odot$ yr$^{-1}$, 
respectively. If we consider that the total mass of the gaseous disk is 
taken as $4 \times 10^{8} \rm M_\odot$, $\sim 1\%$ and $\sim 3\%$ of 
total gaseous disk mass are infalled close to the nucleus within $10^{9}$ yr 
for the models M3 and M4, respectively. Since about half of this gas moves
 into the radius of $0.1$ kpc, the nuclear spirals developed in 
the models M3 and M4 can provide fuels for the nearby AGNs.

\section {DISCUSSION}

\subsection{Effect of Sound Speed in the presence of SMBH }

The effect of the sound speed on the gas flow is somewhat well
understood when there is no SMBH. \citet{eng97}
have noticed that the gas flow in the low sound speed makes the
off-axis principal shocks with a nuclear ring, while that in the
high sound speed ($c_{s} \gtrsim 25$ km s$^{-1}$)
 developes the on-axis ones with no gas on the
$x_{2}$-orbits and hence, this represents the continuous inflow
towards the center. A similar tendency for the principal shocks
to move towards the bar major axis as sound speed increases has been
found by \citet{pat00}. Besides these features, they also found
that the gas inflow inside the radial zone near to the central
regions increases with the sound speed. 

Here we obtained similar results for the effect of the sound speed on 
the gas flow, including the shock locations and the inflow of gas, using 
mass models with and without central SMBHs. However, owing to better
resolutions of our simulations, we can analyze the effect of the sound 
speed on the gas flows in the central kiloparsecs of galaxies where nuclear 
spirals are found to be present. There seems to be a clear distinction
in the morphology of nuclear spirals formed in the central kiloparsecs
according to the presence and absence of the central SMBHs.
In the quasi-steady state, the nuclear regions of the modeled gaseous
disks evolve to the trailing nuclear spirals whose winding angles 
depends on the sound speed in the gas when there is a central SMBH, 
whereas the lower sound speed model that has no SMBH evolves to 
a leading nuclear spiral. However, it should be noted that the 
formation of a leading nuclear spiral such as that of the model M1 is 
limited to some special cases. Even in the models with the same mass model 
as that of the model M1 but with different sound speeds, we do not find 
leading nuclear spirals similar to that found in the model M1. 
The models with higher sound speed ($c_{s} > 15$ km s$^{-1}$) show 
a development of elongated rings, while lower sound speed model 
($c_{s}=5$ km s$^{-1}$) display a double ring structure
of which the thick outer ring is more round and aligned perpendicular to 
the bar and the thin inner ring that is located at the IILR is aligned to 
the bar.

A necessary condition for the formation of a leading nuclear spiral is the 
existence of an IILR but there are other conditions that are required 
for the the formation of a leading nuclear spiral. A favorable condition for the
leading nuclear spiral is the lack of a large amount of gas inflow across
the OILR because
leading nuclear spiral shocks can not propagate outward when there are strong
trailing nuclear spiral shocks from the OILR. In general, 
the leading spiral shocks are generated at the IILR due to the negative
torques of the bar \citep{com02},
while the trailing spiral shocks at the OILR are made
by the orbital switching from the $x_{1}$-orbits to the $x_{2}$-orbits.
In our model M1, the regions around the OILR are devoid of gas particles
after $\sim 5\tau_{bar}$ which weakens the trailing nuclear spiral shocks greatly.
But, there is enough gas near the IILR to make strong leading nuclear spiral
shocks, since gas particles are accumulated at the IILR due to 
 the barrier caused by this resonance. Since the weakening of 
the trailing nuclear spiral shocks in our models is due to the lack of 
gas recycling in our simulations, we expect that strong trailing nuclear spiral 
shocks are more common features than the weak ones in realistic 
physical conditions. This means that the leading nuclear spiral
is mostly a transient feature, if any, that can be formed in some specific
physical conditions. It is the reason why the leading nuclear spirals are so rare
in real galaxies.

The correlation between the winding angle of the trailing nuclear spirals and 
the gas sound speed in the presence of an SMBH that removes the IILR 
can be understood if we consider the dependence of the angular momentum
evolution on the sound speed in the gas (see Fig. 10). The large winding angles 
of nuclear spirals in the higher sound speed models are due to 
the large amount of angular momentum loss which makes physical condition 
plausible for the innermost parts of nuclear spirals to reach very 
close to the center. On the other hand, the gas particles in 
the lower sound speed models lose their angular momentum slowly so that
they can not spiral in close to the center. This interpretation is also
supported by the correlation between the gas inflow in the very vicinity of the 
nucleus ($r \sim 0.2$ kpc) and the sound speed in the gas (see  Fig. 9).
The quasi-steady state morphology of the trailing nuclear
spirals are consistent with the prediction of the linear density wave
theory in the sense that the spiral arms open out as sound speed in the
 gas increases \citep{eng00}.

\subsection {Hydrodynamic Shock as an Origin of Nuclear Spirals}

The gas density waves proposed by \citet{eng00} is one of promising 
mechanism that can explain the formation of the grand-design nuclear spirals.
But, it can be applicable only to the low amplitude spirals where no 
shocks exist because gas density waves can not propagate in the
regions where shocks are strong enough to drive non-linear response of
gas flows. There are several examples of grand-design nuclear spirals which 
have densities high enough for active star formation. The Hubble Space
Telescope $NICMOS$ $H$-band images of NGC 5427 and NGC 5614 clearly
show young stellar populations in their nuclear spirals \citep{mar03}.
Since the densities of their nuclear spirals are too high to be considered 
as linear, they can not be explained by the gas density waves of \citet{eng00}
but by the hydrodynamic spiral shocks presented in this study. 
As shown in Fig. 7, the spiral shocks can penetrate deep inside the ILR,
at least in early phase of evolutions (t $\lesssim 5 \tau_{bar}$) to
wind upto $2 \pi$ radian in azimuthal angle 
although there is a certain drop of particle density at 
$\sim {3 \over 4} \pi$. 
The shock nature of the nuclear spiral is also addressed by \citet{mac02}
from the grid-based hydrodynamical simulations. In their high sound speed
run ($c_{s}=20$ km s$^{-1}$), they found that a two-armed nuclear spiral 
is a direct continuation of the principal shock along the leading edge of 
the bar and it reaches close to the nucleus well beyond the IILR
 where the gas density waves can not penetrate. 
In their model, the strength of shocks, as measured by div$^{2} {\bf v}$,  
and the arm/interarm density contrast are comparable to those of the
principal shocks, which indicates that such a nuclear spiral is well beyond
the linear regime that can be explored in the density wave theory.

However, at the late phase of evolution when the gas flows become a
quasi-steady state, we do not expect strong shocks deep inside the ILR
(or the OILR when there are two ILRs) due to less orbital crossings, since 
most of the particles are already in the x2-orbits.
Thus, the maintenance of nuclear spiral in the quasi-steady state is 
not made by the bar-driven spiral shocks but by the gas density waves proposed 
by \citet{eng00}. But, since the weakening of the nuclear 
spiral shocks is due to the devoid of gas in regions of the principal 
shock, the nuclear spiral turns into density wave only when the driver 
of the nuclear spiral shocks ceases. 
Of course it is also possible that grand-design nuclear spirals
 can be made by gas density waves. But, nuclear spirals formed by
gas density waves should be of low-amplitude so that the gas response
could be considered as a linear wave \citep{eng00}. 
It is worth to note that gas density wave itself is triggered by 
 the spiral shocks driven by the bar \citep{eng00}.

The self-gravity of gas is known to be effective to drive gas inflow
by generating torques among gas clumps \citep{wad92,elm94,fuk98,fuk00}.
However, the density of the gaseous disks should be high enough 
to make clumps by the gravitational instability.  
To clarify whether the inflow of the gas towards the
center is due to the hydrodynamic
shocks or due to the torques generated by the
gas clumps, we calculated Toomre's $Q$-value in
the several high density regions where the spiral arms are developed.
If the value of the $Q \lesssim 1$, the gaseous disk would be
gravitationally unstable, whereas  $Q > 1$ represents the stable gaseous disk.
Toomre's $Q$-value of our models is expressed as 
\begin{eqnarray}
Q & \sim & 3.44 \times 10^{-3} \frac {\kappa} {\sigma}
\left(\frac {c_{s}} {10 km s^{-1}}\right),
\end{eqnarray}
where $c_{s}$ is the sound speed corresponding to each model,
$\sigma$ is the surface density of the selected high density regions of
the gaseous disk at a particular evolution time and $\kappa$ is the
epicycle frequency at the location of each selected region.
Here the units of the $\sigma$ and $\kappa$ are
$2 \times 10^{5} \rm M_{\odot}$ pc$^{-2}$ and
$929.5$ km s$^{-1}$ kpc$^{-1}$, respectively. We found that the
surface density $\sigma$ of all the selected regions are found to be
smaller than the term $3.44 \times 10^{-3} \kappa \left({c_{s}} /
{10 km s^{-1}}\right)$ of Eq. $(17)$ and as a result, Toomre's $Q$-values 
are greater than unity (i.e., $Q >1$) for our models.
The reason for the low surface density even at the densest regions of the 
gaseous disk is due to the small mass fraction of the initial gaseous
disk, i.e., $M_{gas}/M_{G}=0.01$.
Although some clumped structures seem to appear at later phase of evolutions
for the model M4, it is clear that the self-gravity of the gas is not
large enough to drive  the gas particles inward.
Thus, we confirm that the inflow of the gas
particles towards the center to form the symmetric two-armed nuclear
spiral is due to the hydrodynamic shocks induced by the gravitational
torque of the bar rather than the torques among the gaseous clumps. 
Despite this fact, the gas self-gravity still takes its importance to include
in the simulations, as it can affect the nuclear gaseous structure 
in the late phase of evolutions that we have already noticed 
in the case of models M1 and M4.

\section{CONCLUSIONS}

We have used the SPH technique to simulate the response of the
gaseous disk to the moderately strong bar potentials in order to
understand the formation mechanism of the symmetric two-armed nuclear
spirals. We found that the gas flows driven by the gravitational torques 
of the bar lead to the formation of nuclear spirals whose morphology 
depends much on the sound speed of gas. 
Although SMBHs play a critical role in shaping
the gravitational potential in the central kiloparsec to weaken or to remove
the IILR that prevents the gas inflow close to the nucleus,
the sound speed of gas seems to be the primary physical parameter
which controls the gas flows deep inside the ILR to make nuclear features.
In the quasi-steady state, the global morphological features of 
the modeled gaseous disks are not much affected by the sound speed 
in the gas although models with high sound speeds evolve faster than 
those with lower sound speeds. Since mass inflow from the disk is an
essential process to make nuclear features, proper treatment of 
the global evolution of the gaseous disk is required to understand 
the nuclear features such as nuclear spirals. 

The symmetric two-armed nuclear spirals are
formed by the hydrodynamic spiral shocks induced by 
the gravitational torque of the bar when there is 
a central SMBH. The masses of SMBHs are 
assumed to be about $1\%$ of the total visible mass of 
the model galaxies that is massive enough to remove the IILR. 
The symmetric two-armed nuclear spirals 
whose innermost parts reach close to the center of galaxies are
likely to form in the high sound speed medium
($c_{s} \gtrsim 15$ km s$^{-1}$), while the ring-like nuclear
spirals develop in the low sound speed medium. In some special physical 
conditions like those assumed in the cold gas models
of present study, a  leading nuclear spiral can be developed between 
the IILR and the OILR when there is no central SMBH. In this case, there is 
little inflow of gas inside the IILR due to IILR barrier. In the presence of 
a central SMBH, the mean gas inflow rate inside the
central hundred parsecs increases with the sound speed. For the model that
assumes sound speed of $c_{s}=20$ km s$^{-1}$, the mean gas inflow rate
within $0.1$ kpc is about $\sim 0.007$ $\rm M_\odot$ yr$^{-1}$, 
which suggests that $\sim 1.75 \%$ of the total gaseous disk mass moves 
into the radius of $0.1$ kpc within $10^{9}$ yr. Thus, the nuclear spirals
formed in the hot interstellar medium seem to be effective for 
fueling the nearby AGNs. Finally, the trailing nuclear spiral shocks are too 
weak to support the nuclear spiral in the later phase of evolution
when there is little gas inflow across the ILR. Thus, nuclear spiral turns 
into density wave when the driver of spiral shocks ceases and 
the spiral arms open out as sound speed in the gas increases.

\acknowledgments {We thank the anonymous referee for useful remarks and 
suggestions which improve the present paper
enormously. We also wish to express our sincere thanks to 
Dr. Roger Fux who provides the PMSPH code. HBA  thanks 
Prof. Hyesung Kang for valuable discussion and comments on the numerical
simulations. PT would like
to express his sincere thanks to ARCSEC (Astrophysical Research Center
for the Structure and Evolution of the Cosmos) for providing support
through the postdoctoral fellowship which made this study possible.
This work is supported in part by grant No. R01-1999-00023
from the Korea Science Engineering Foundation (KOSEF).
Most of the computations are conducted using the super computer
facilities in the Korea Institute of Science and Technology Information
(KISTI).}

\clearpage

\begin{figure}
\plotone{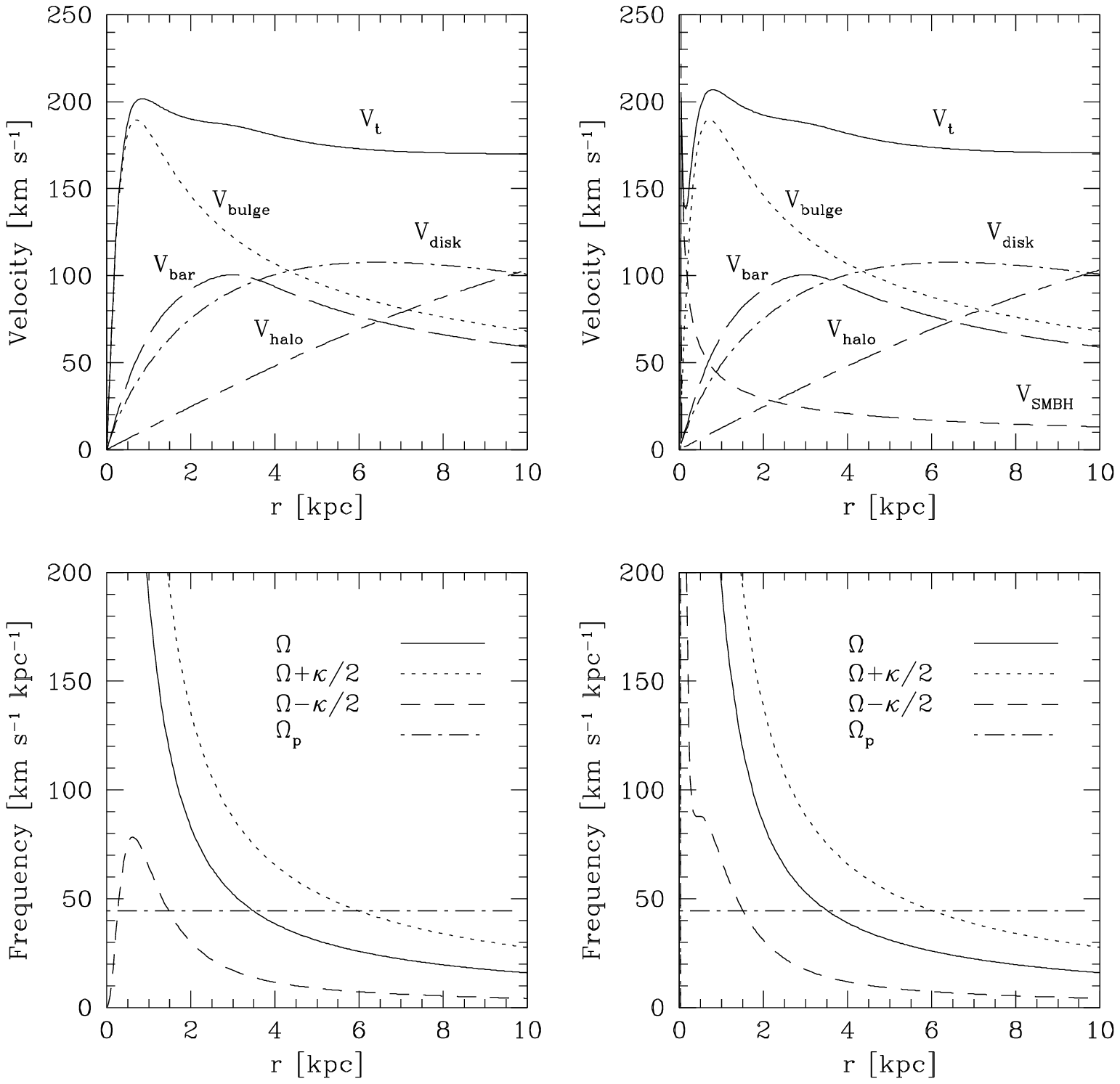}
\caption{Rotation curves and angular frequencies of model
galaxies. Because of non-axisymmetric nature of the bar, its
 contribution is included here after averaging it axisymmetrically.
 The left panels are for
the model M1 that has no SMBH, whereas the right panels are for the
 model M2 with SMBH whose mass is about $1$ \% of the total visible
 mass of galaxy. The horizontal dot-dash lines in the bottom panels 
represent the  bar pattern speed of 
$\Omega_{p}$=44.6 km s$^{-1}$ kpc$^{-1}$. }
\end{figure}
\clearpage
\begin{figure}
\plotone{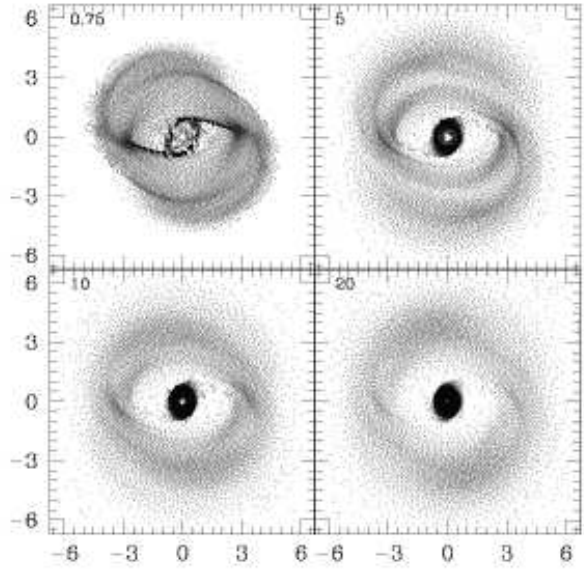}
\caption{Global evolution of gaseous disk for the model M3. The number
 in the top left corner of each panel is the
 evolution time in the unit of $\tau_{bar} \sim 1.4 \times 10^{8}$ yr. 
 The bar lies horizontally and the box size is $13.4$ kpc in one dimension.}
\end{figure}
\clearpage
\begin{figure}
\plotone{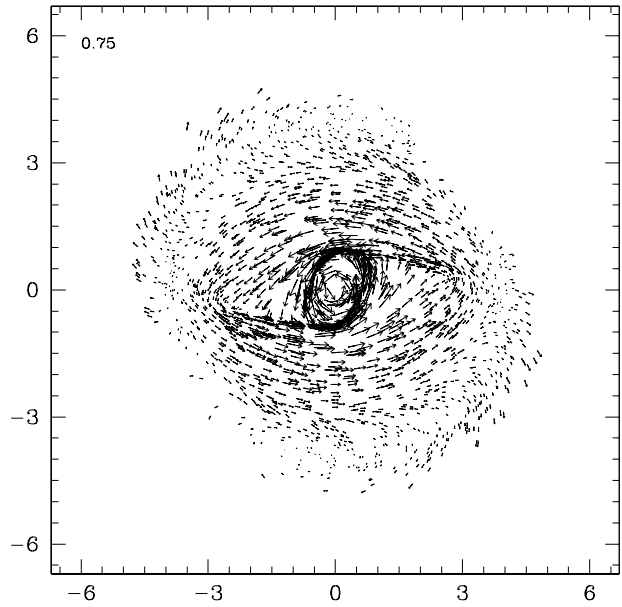}
\caption{Velocity field of the Model M3 at evolution time of 
$0.75 \tau_{bar}$. The velocity of each particle is 
represented by the arrow whose length is proportional to the
speed. To achieve better resolution, the velocities of randomly
selected $2 \times 10^{3}$ gas particles are plotted. The bar 
lies horizontally and the box size is $13.4$ kpc in one dimension.}
\end{figure}
\clearpage
\begin{figure}
\plotone{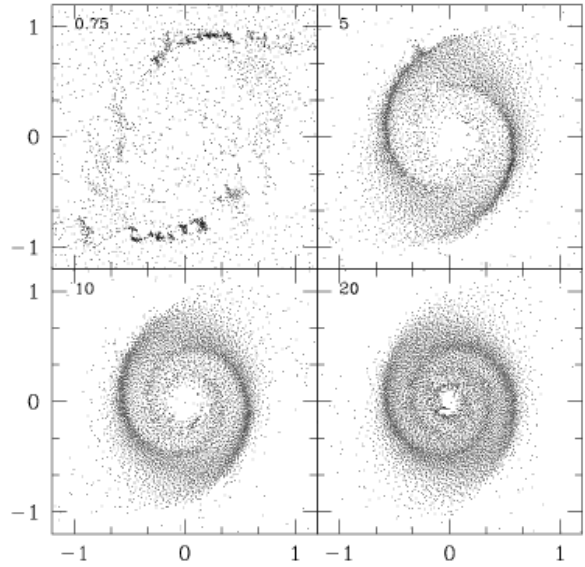}
\figcaption{Evolution of the nuclear regions of the gaseous disk 
for the model M3. The number in the top left corner of each panel 
is the evolution time in the unit of $\tau_{bar}$. The bar lies horizontally 
and the box size is $2.4$ kpc in one dimension.}

\end{figure}
\clearpage
\begin{figure}
\plotone{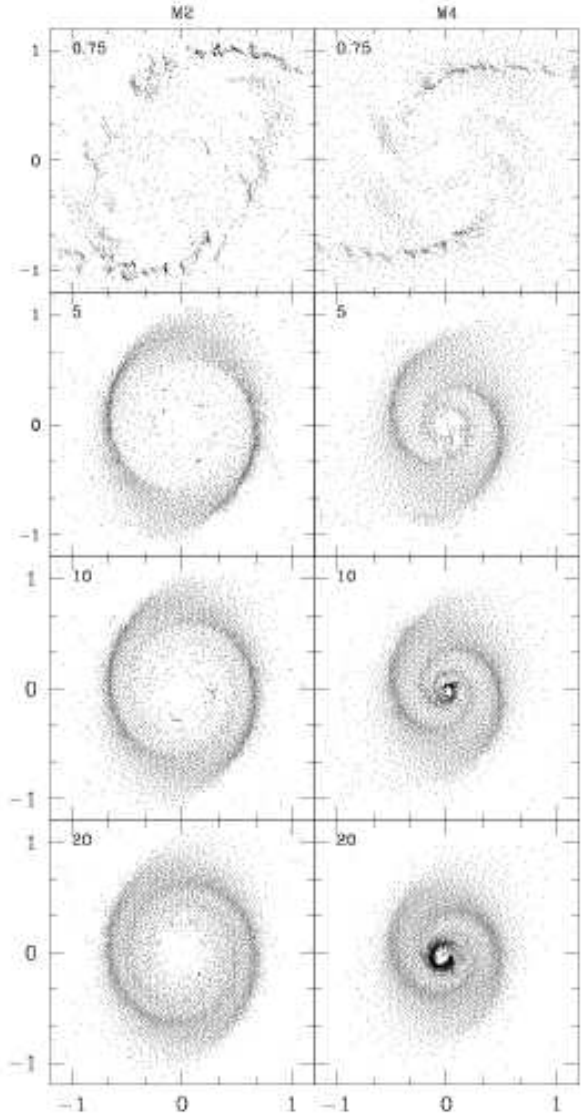}
\caption{Same as Fig. $4$, but for the models M2 and M4. The left panels
 are for the model M2, whereas the right panels are for the model M4.}
\end{figure}
\clearpage
\begin{figure}
\plotone{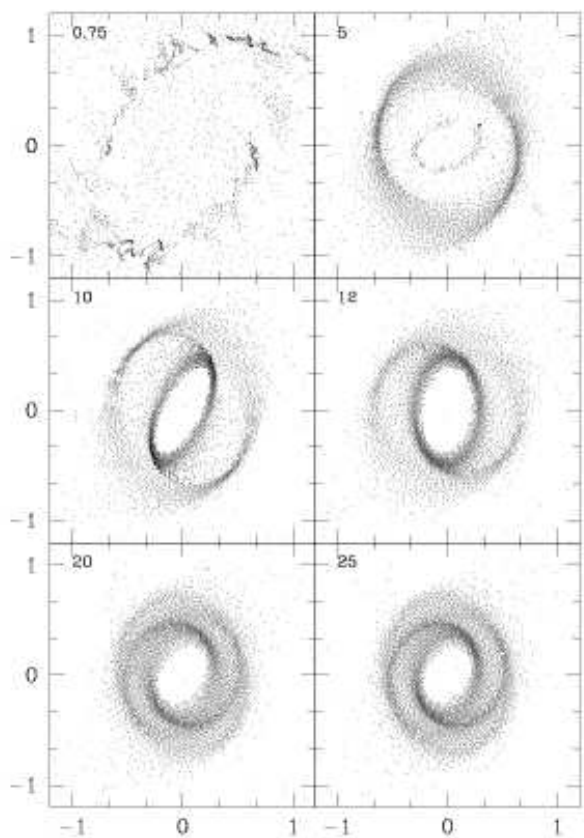}
\caption{Same as Fig. $4$, but for the model M1.}
\end{figure}
\clearpage
\begin{figure}
\plotone{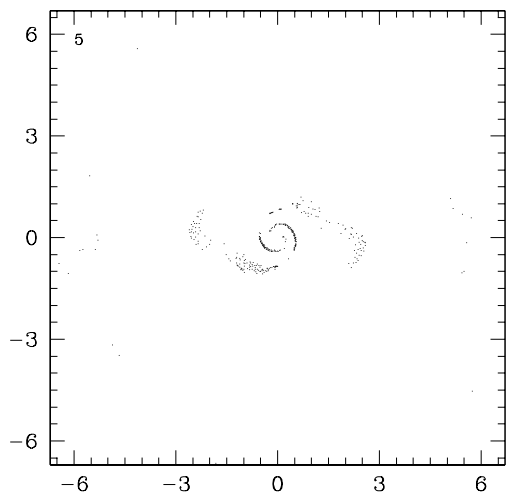}
\caption{Distribution of the gas particles that are shocked
by supersonic flow (i.e., $-h {\bf \nabla} \cdot {\bf v} > c_{s}$)
for the model M3 at  evolution time of $5 \tau_{bar}$. The bar lies 
horizontally and the box size is $13.4$ kpc 
in one dimension.}
\end{figure}
\clearpage
\begin{figure}
\plotone{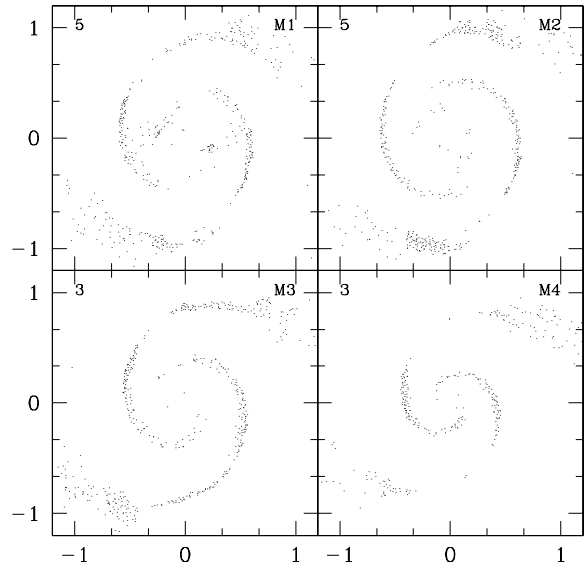}
\caption{Nuclear distribution of the gas particles that are shocked
by supersonic flow (i.e., $-h {\bf \nabla} \cdot {\bf v} > c_{s}$)
for the models M1-M4. The models are indicated  in
the top right corner and the evolution time in the unit of
$\tau_{bar}$ is given in the top left corner of each panel. 
The bar lies horizontally and the box size is $2.4$ kpc 
in one dimension.}
\end{figure}
\clearpage
\begin{figure}
\plotone{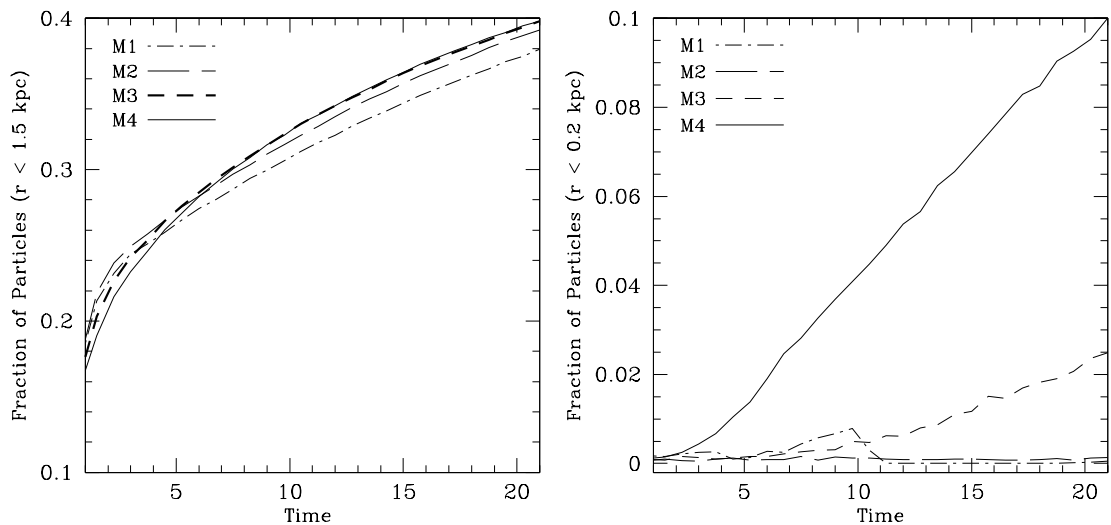}
\caption{The inflow of the gas particle in the models M1-M4
for two different radial zones. The time evolution of the fraction
of particles accumulated inside the radius of $1.5$ kpc is
presented in the left panel, while that for the
radius of $0.2$ kpc is shown in the right panel. The
time is given in the unit of $\tau_{bar}$. The line styles
corresponding to each model are designated in the top left corner
of each panel.}
\end{figure}
\begin{figure}
\plotone{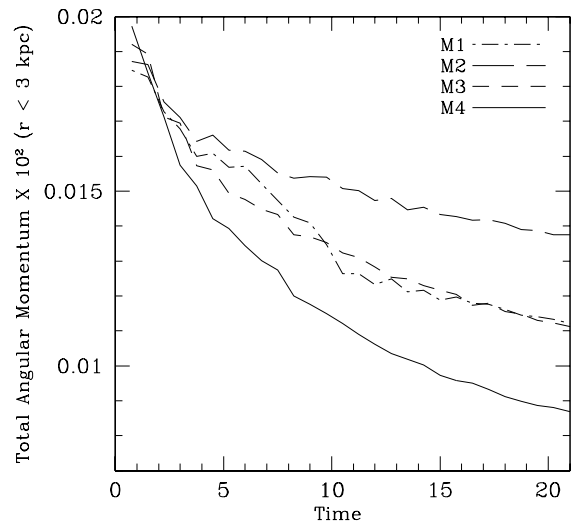}
\caption{The time evolution of the total angular momentum of the
gas particles accumulated  within  the radius of $3$ kpc
(i.e., the bar radius) for the models M1-M4. The
evolution time is given in the unit of
$\tau_{bar}$. The line styles
corresponding to each model are given in the top right
corner of the panel.}
\end{figure}
\end{document}